\documentclass[manuscript]{acmart}

\AtBeginDocument{%
  }

\setcopyright{acmlicensed}
\copyrightyear{2024}
\acmYear{2024}

\acmConference[XAIxArts 2024]{Explainable AI for the Arts Workshop 2024}{June 23, 2024}{Chicago, IL, United States}
\acmISBN{978-1-4503-XXXX-X/18/06}

\begin{document}

\title{Embodied Exploration of Latent Spaces and Explainable AI}

\author{Elizabeth Wilson}
\email{e.j.wilson@arts.ac.uk}
\orcid{0000-0001-6212-6627}
\affiliation{%
  \institution{Creative Computing Institute, University of the Arts London }
  \city{London}
  \country{UK}
   \postcode{SE5 8UF}
}

\author{Deva Schubert}
\email{deva.schubert@gmx.de}
\orcid{}
\affiliation{%
  \institution{Independent Artist}
    \city{Berlin}
    \country{DE}
}

\author{Mika Satomi}
\email{mika@kobakant.at}
\orcid{}
\affiliation{%
  \institution{KOBAKANT}
   \city{Berlin}
    \country{DE}
}

\author{Alex McLean}
\email{alex@slab.org}
\orcid{}
\affiliation{%
  \institution{Then Try This, Cornwall}
    \country{UK}
}

\author{Juan Felipe Amaya Gonzalez}
\email{info@jfag.co}
\orcid{}
\affiliation{%
  \institution{Independent Artist}
   \city{Berlin}
    \country{DE}
}

\renewcommand{\shortauthors}{Wilson et al.}

\begin{abstract}

In this paper, we explore how performers' embodied interactions with a Neural Audio Synthesis model allow the exploration of the latent space of such a model, mediated through movements sensed by e-textiles. We provide background and context for the performance, highlighting the potential of embodied practices to contribute to developing explainable AI systems. By integrating various artistic domains with explainable AI principles, our interdisciplinary exploration contributes to the discourse on art, embodiment, and AI, offering insights into intuitive approaches found through bodily expression.

\end{abstract}



\keywords{Musical AI, Explainable AI, E-textiles, Live Coding}


\maketitle

\section{Introduction}

In this position paper, explainable AI is examined through embodied latent space exploration, referencing a performance at Transmediale Studios, Berlin in March 2023. Details of creating an AI-informed artistic performance are provided, challenging conventional notions of intelligence and exploring AI's esoteric and social dimensions through rituals. The performance connected the AI model `RAVE' (Real-Time Audio Variational autoEncoder \cite{caillon2021rave}) to three separate artistic practices: e-textiles, live coding, and performance art. Through their interplay, the exploration of the latent space of the model trained on the performers' vocal input became apparent, where the model became material, and bodies in space were the main vectors with which to explore potential possibilities offered by the latent space. In terms of AI concepts, this piece aimed to challenge mainstream ideas of AI, such as the emulation of human neural
pathways as the method of creating new intelligences. A collective view of intelligence was given as the impetus for the performance: by treating intelligence as collaborative, the aim was to undermine the dualist thinking of body/mind as well as science/fiction,
and open ourselves to a more human-centric approach to AI, in terms of pattern generation, pattern recognition and
connected-ness, but also resonance and textility.

\section{xAI for the Arts}

The field of eXplainable AI (XAI) investigates methods to enhance the comprehensibility of machine learning models for individuals, thereby expanding their practicality and enabling non-specialists to employ them across diverse scenarios. Specifically, XAI experts delve into strategies for understanding complex and opaque AI models, like neural networks and deep learning methods, which are often challenging for laypeople to grasp.

Systems that are easy to use and understandable are desirable to artists for many obvious reasons, such as the need to avoid overtly steep learning curves or opaque technical processes. 
\citet{bryan2023proceedings} were among the first to directly tackle the issue of eXplainable Artificial Intelligence (XAI) in the arts, with some particular works focused on music (e.g. \cite{llano2020}, \cite{bryan2024exploring}) are applicable to this work. Overall, within the community the ideas that AI models may be exploited as both an artwork and a form of explanation is not uncommon. The latent space often provides artistic impetus, sometimes providing the not-yet-conceptualised to artists. Explainability becomes embedded where the navigating latent space can be used to explain its limits \cite{bryan2024exploring}.

\section{The Context}

E-Textiles are fabrics that integrate electronic components like batteries, lights, sensors, and micro-controllers. For this performance, e-textiles were used to create soft and flexible sensors that were worn on the body to sense movements. Pressure sensors were explored  \cite{luo2023technology} made of Eeontex stretch resistive fabric to sense bend and
stretch of garments caused by body movements. The Bela mini
board was used which allowed us to read up to 8 analogue sensors per performer and 1) directly trigger sound from a single sensor
data, 2) process multiple sensor data with machine learning to be used as a reduced number of synthesis parameters.
It also used ml.lib externals on Pure Data open source software \cite{bullock2015ml}, classifying specific gestures and enabling the system to recognise repetitions in the performers' movements.

Live coding, in this context, refers to artistic programming
onstage as performance with dedicated systems, often in part or wholly improvised \citep{blackwell2022live}.
TidalCycles was the main software used for live coding sounds \cite{mclean2014making}, however a few custom elements were
employed to alter the territories that live coders were used to, pushing them into new modes of interacting with sound
through software. The two live coders were tasked with working independently when manipulating sensorial information from the materials but also collaboratively in response to each other.

\begin{figure}[h]
  \centering
  \includegraphics[width=0.35\linewidth]{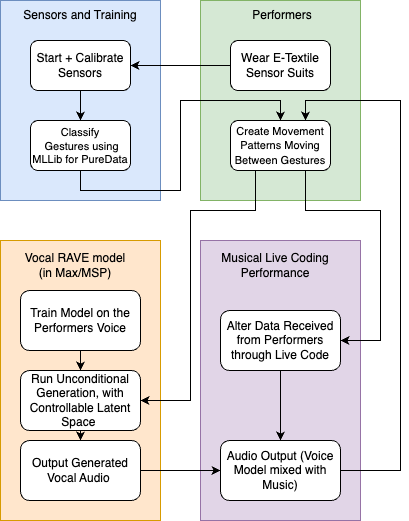}
  \caption{}
  \Description{}
  \label{fig:overview}
\end{figure}

One artistic use of AI in the performance was to create an artificial voice that could be explored with both live coding and the sensor information gathered from the textiles.
To create the sounds from an artificial voice, the Neural Audio Synthesis model RAVE \cite{caillon2021rave} was employed. In short, this uses a Variational Auto Encoder (VAE) structure but is applied as a real-time
model (able to generate 48kHz audio signals, while simultaneously running 20 times faster than real-time on a standard
laptop CPU). The RAVE-model is trained on a corpus of audio generated by the performers voices, and the VAE structure decodes and then encodes the
signals. The encoding and decoding can be controlled internally from the MaxMSP software, or via external messaging protocols. Data from the sensors on the face of the four performers was sent using the zero-mq (zmq)
protocol \citep{hintjens2013zeromq}. An overview detailing  the specific interactions between the performers and the AI model can be seen in figure \ref{fig:overview}.

\section{The Performance}

The performance took place in March 2023 at Transmediale Studios, in Berlin, DE. It was designed as a durational performance, lasting four hours, with visitors invited to come and go throughout. 
It was an inter-disciplinary collaboration between two live coders (who use computer code as the means for artistic expression \cite{nilson2007live}), five performers, and one e-textile artist.

\begin{figure}[h]
  \centering
  \includegraphics[width=0.75\linewidth]{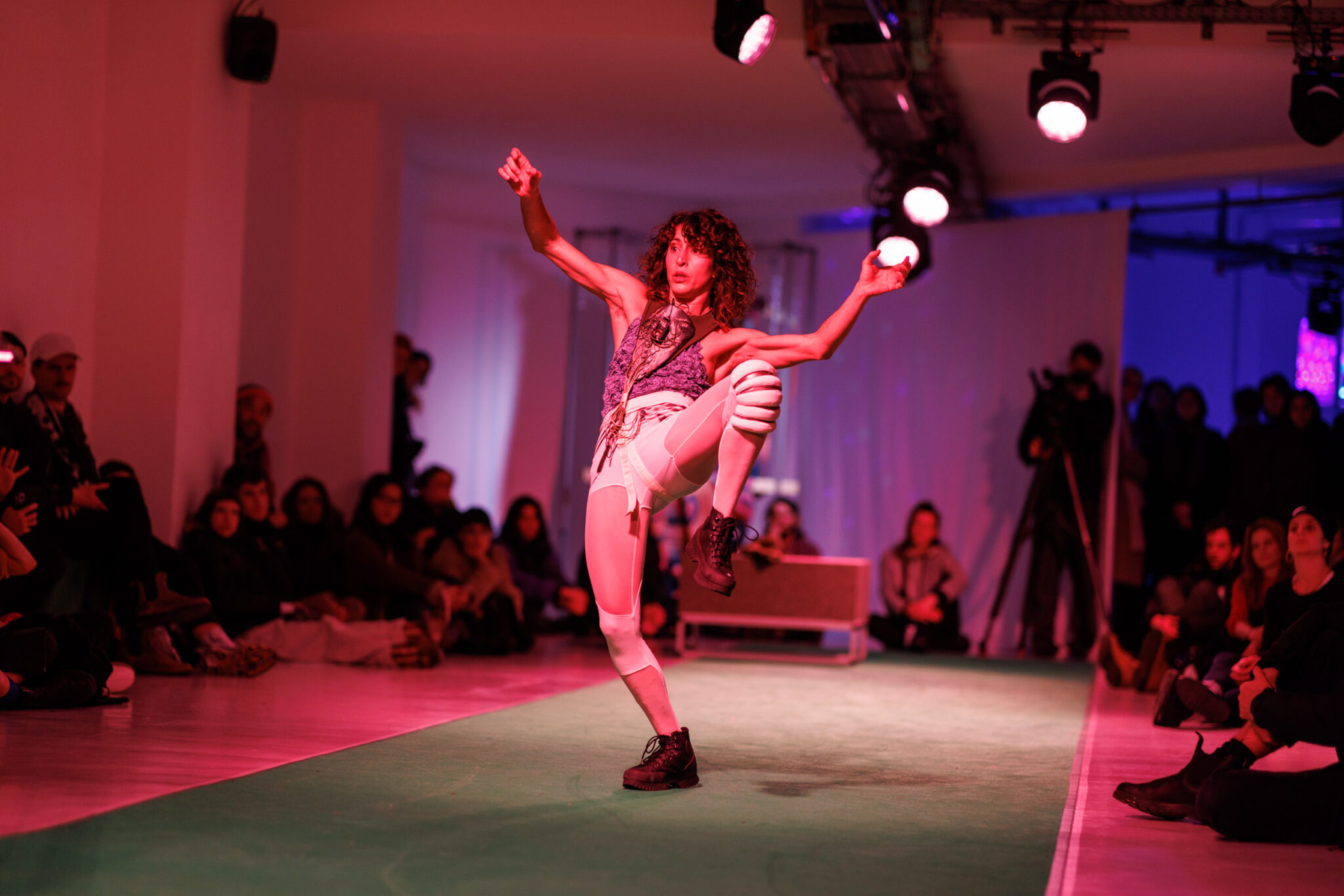}
  \caption{Performer Manoela Rangel exploring latent decoding with sensors on the body. Image Credit: Frank Sperling, \url{https://franksperling.net/Info}.}
  \Description{A performer with a leg bent in the air, wearing white pants with sensors embedded}
\end{figure}

A video summary of the performance can be found at
\url{https://vimeo.com/882219756}.

\section{Reflections}

An interesting case for explainable AI emerged through developing this performance. The aim was to integrate AI into artistic practice, but experimentation revealed that performers' bodies were crucial for understanding. The performers engaged in a tactile dialogue with the algorithms, with their movements influencing the traversal of the latent space. Repetitive movements helped explore the voice model's outcomes, and performers often wanted to relate the sounds of the voice to their movements, especially when phrases or words became more intelligible.

VAE models are not inherently deterministic. In a traditional autoencoder, the encoder network maps the input data to a fixed-length latent vector, resulting in a deterministic encoding process. However, in a VAE model, the encoder network maps the input data to a probability distribution in the latent space, typically Gaussian. This means that instead of producing a single fixed-length vector, the encoder outputs parameters (mean and variance) of a distribution from which latent vectors are sampled. During training, the VAE models learns to encode input data into the latent space such that similar data points are mapped close together in the latent space. Sampling from the learned latent space during inference introduces stochasticity, as each sample drawn from the latent space will generate different output data when decoded by the decoder network \cite{doersch2016tutorial}. This property of the VAE model allowed connections between bodily expression and new ways of learning about the intricacies of the system itself. Where the focus of the piece was on integrating AI into artistic practice, the performers' interactions with their bodies unexpectedly emerged as a pivotal aspect of the exploration. Through this embodied interaction, the performers not only contributed to the artistic expression of the piece but also helped provide a use case for the body as a means with which higher-dimensional spaces can be explored. This unexpected synergy between performers' bodies and AI technology highlights the potential for embodied practices to inform and enrich the development of explainable AI systems.

\begin{acks}
This work was supported by the Volkswagen Foundation LINK masters Grant and the Fonds Darstellende Künste Foundation Grant. Additionally, Wilson’s contributions were supported by EPSRC and AHRC under the EP/L01632X/1
(Centre for Doctoral Training in Media and Arts Technology) grant and McLean’s contributions were supported by a
UKRI Future Leaders Fellowship [grant number MR/V025260/1].
\end{acks}

\bibliographystyle{ACM-Reference-Format}

\end{document}